\documentclass[11pt]{article}
\usepackage{fleqn,cospar}
\usepackage{graphicx}

\title{MM/SUBMM OBSERVATIONS OF SYMBIOTIC BINARY STARS:
IMPLICATIONS FOR THE MASS LOSS AND MASS EXCHANGE\footnote{Paper presented at COSPAR 2000 ``New Results in FIR 
and Submm Astronomy'', to be published in Adv. Space Res.}}

\author{J. Miko{\l}ajewska\address{Copernicus Astronomical Center, 
      Bartycka 18, PL-00716 Warsaw, Poland},
      R.J. Ivison\address{Department of Physics \& Astronomy, 
      University Colleage London, Gower Street, London WC1E 6BT}
      and
      A. Omont\address{Institut d'Astrophysique, CNRS, 98 bis Bvd Arago,
      75014 Paris, France}}

\begin{document}

\maketitle

\begin{abstract}
We discuss mm/submm spectra of a sample of symbiotic binary systems,
and compare them with popular models proposed to account for their radio
emission. We find that radio emission from quiescent S-type systems
orginates from a conical region of the red giant wind ionized by
the hot companion (the STB model),
whereas more complicated models involving winds from both components and their
interaction are required to account for radio emission of active systems. 
We also find that the giant mass-loss rates 
derived from our observations are systematically higher than those
reported for single cool giants. This result is in agreement with 
conclusions derived from IRAS observations and with requirements of models
for the hot components.
\end{abstract}

\section*{INTRODUCTION}

Symbiotic stars are long-period interacting binaries composed
of an evolved cool giant and a hot and luminous companion --
in most cases a wind-accreting post-AGB star -- surrounded 
by an ionized nebula. 
The nature of the cool giant plays a key role in the symbiotic 
phenomenon: it constrains the size of the binary, which must
have enough room for a red giant, and yet allow the giant to
transfer sufficient mass to its companion. As a result, we have 
two distinct classes of symbiotic binaries: the S-type (stellar)
with normal red giants and orbital periods of about 1--15 yr,
and the D-type (dusty) with Mira primaries usually surrounded
by a warm dust shell and orbital periods generally longer than 
15 yr.
The hot star in most cases appears to be a white dwarf powered
by thermonuclear burning of the material accreted from its 
companion wind.
The presence of both the evolved giant, often heavily losing mass,
and the hot companion copious in ionizing photons and in many cases 
possesing its own wind lends a large variety to the circumstellar
environment of symbiotic stars.
In particular, one can expect: ionized and neutral region;
dust forming region(s); accretion/excretion disks; interacting
winds; bipolar outflows and jets.

Such a complex multi-component structure makes symbiotic stars
a very attractive laboratory to study evolution and interaction 
of binary systems, and excellent targets for both ground-based
and space observations in practically any spectral range.
The radio and far infrared studies are here of great importance as
they can best probe the circumstellar evelopes of symbiotic 
binaries providing information about stellar winds and their interaction. 

\begin{figure}[t]
\begin{center}
\begin{minipage}[t]{80mm}
\includegraphics[width=80mm]{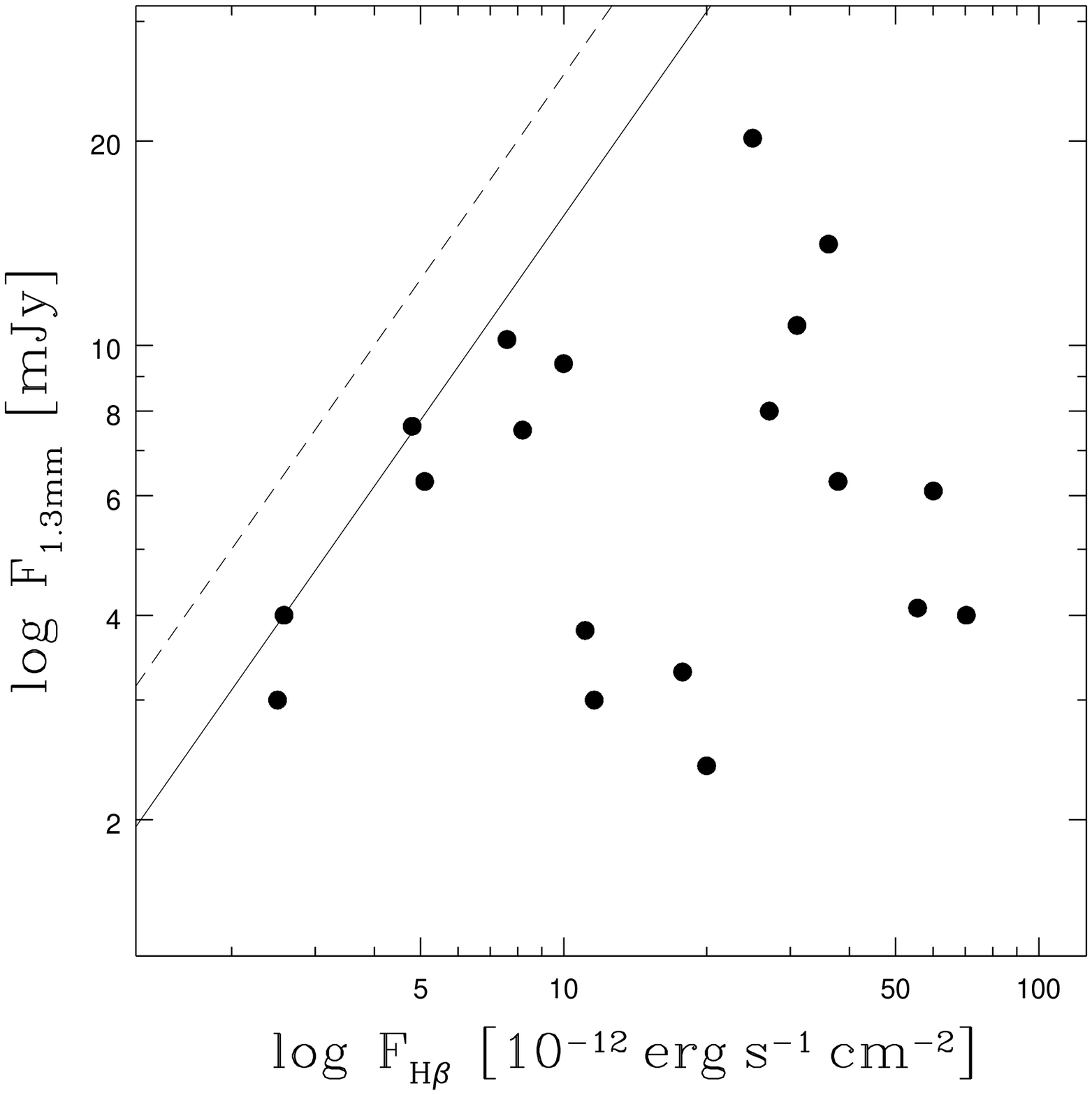}
\par\vspace{0pt}
\caption{\label{fig:fig1} The 1.3\,mm flux density vs.
the H$\beta$ flux corrected for interstellar extinction.
The solid and dashed lines correspond to optically thin case B emission 
for $T_e = 10^4$ and $2\times10^4$ K, respectively.}
\end{minipage}
\hfill
\begin{minipage}[t]{80mm}
\includegraphics[width=80mm]{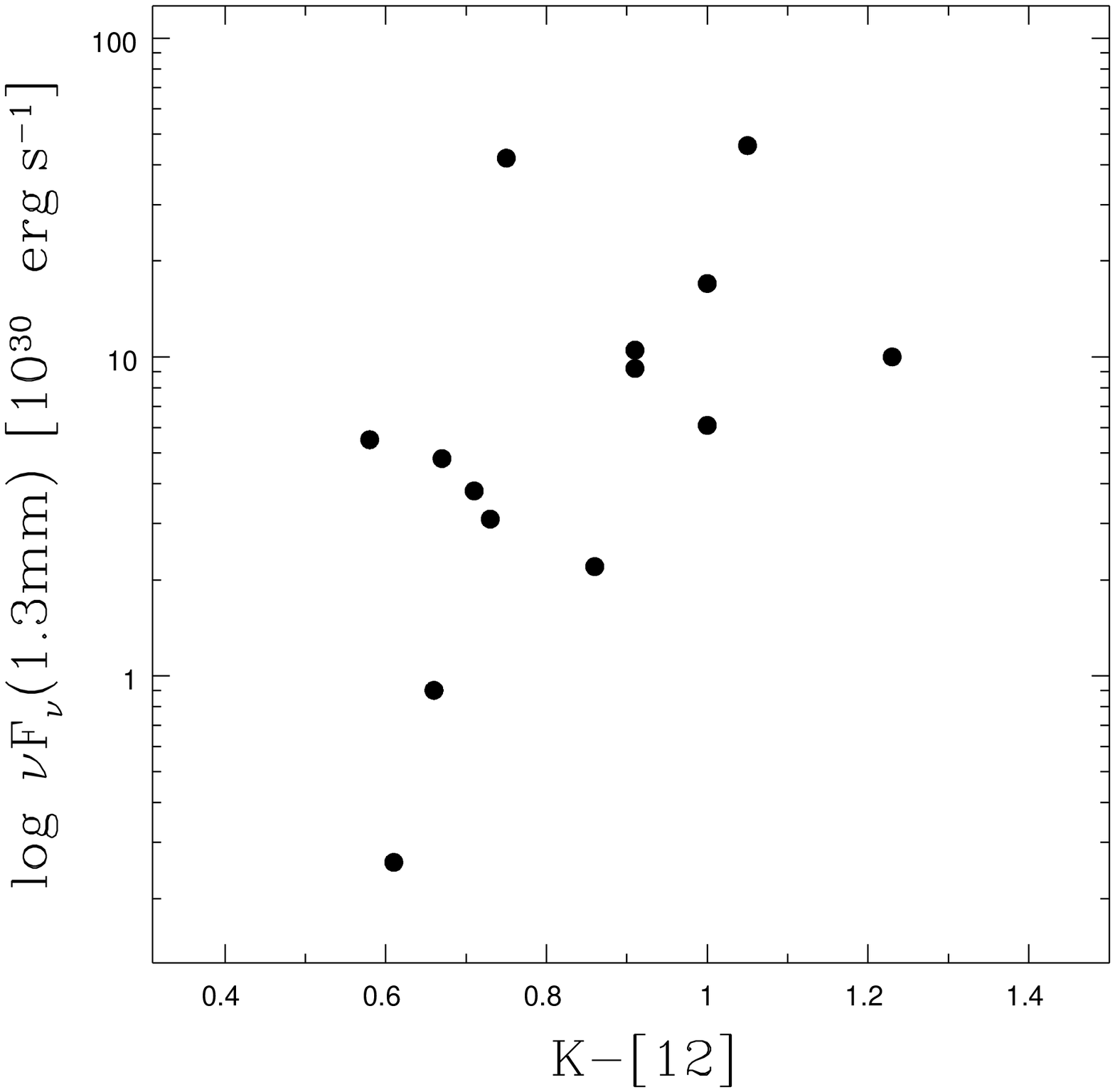}
\par\vspace{0pt}
\caption{\label{fig:fig.2} The radio luminosity vs. $K-[12]$ colour , which is
 believed to measure the circumstellar dust emission for cool giants.}
\end{minipage}
\end{center}
\end{figure}

So far $\sim 25$\% of all symbiotic stars have been detected at cm radio range. 
In practically all cases, the radio emission is consistent with free-free 
radiation from ionised gas (Seaquist \& Taylor 1990, hereafter ST90). Seaquist 
et al. (1984) proposed a simple binary model (the STB model) in which the radio 
emission originates from a conical region of the red giant wind ionized by the 
hot companion. The geometry of this region is governed by a single parameter, 
$X$, which depends on the red giant mass-loss rate, the binary separation, and 
the Lyman continuum luminosity of the ionizing source. The predicted radio 
spectrum turns over from optically thick to optically thin emission at a 
turnover frequency, $\nu_{\rm t}$, which is related to the binary parameters. 
The observations of  $\nu_{\rm t}$ in quiescent S-type systems with known 
orbital parameters thus provides a critical test for the STB model. 
Unfortunately, the spectral turnovers have been thusfar determined only in 
either D-type systems (where orbital periods are not known) or S-type systems --
 e.g. AG Peg -- recovering from nova-type outbursts (Seaquist \& Taylor 1992; 
Ivison et al. 1992, 1995).

In the following, we present and discuss preliminary results of mm/submm 
observations of a sample of quiescent S-type symbiotic systems. In particular, 
we show that they are in general consistent with the predictions of  the STB 
model. We also estimate mass-loss rates for the cool giant and discuss 
relations between intensity of the mass loss and other parameters of these 
binary systems.

\section*{OBSERVATIONS}

In February 1997, a sample of about 40 symbiotic stars was surveyed at 1.3 mm 
witth the IRAM 30-m MRT. In 1997-98, 12 of these objects were also observed 
with the JMCT at 2, 1.3, 0.85 and 0.45 mm, respectively. We have combined our 
mm/submm flux densities with other published data to provide information about 
the continuum spectra in the radio/IR region for all of the symbiotic systems 
in our sample.  In particular, cm-wave data were from Seaquist \& Taylor (1990, 
hereafter ST90) and Seaquist et al. (1993, hereafter SKT93); $IRAS$ fluxes were 
from Kenyon et al. (1988); near-IR  photometric data were from Kenyon (1988) 
and Munari et al. (1992). We have ignored the possible effect of time 
variability on the interpretation of these spectra although we are aware of its 
importance, at least in some systems. 

For the following analyses we have selected 20 S-type systems quiescent at the 
time of our observations. Most of these systems are also well-studied in the 
optical and UV; the appropriate binary periods are known for 14 of them, and 
for 7 systems we also have spectroscopic orbits (Belczy{\'n}ski et al. 2000). 
symbiotic binaries.

\begin{figure}[t]
\begin{center}
\begin{minipage}[t]{80mm}
\includegraphics[width=80mm]{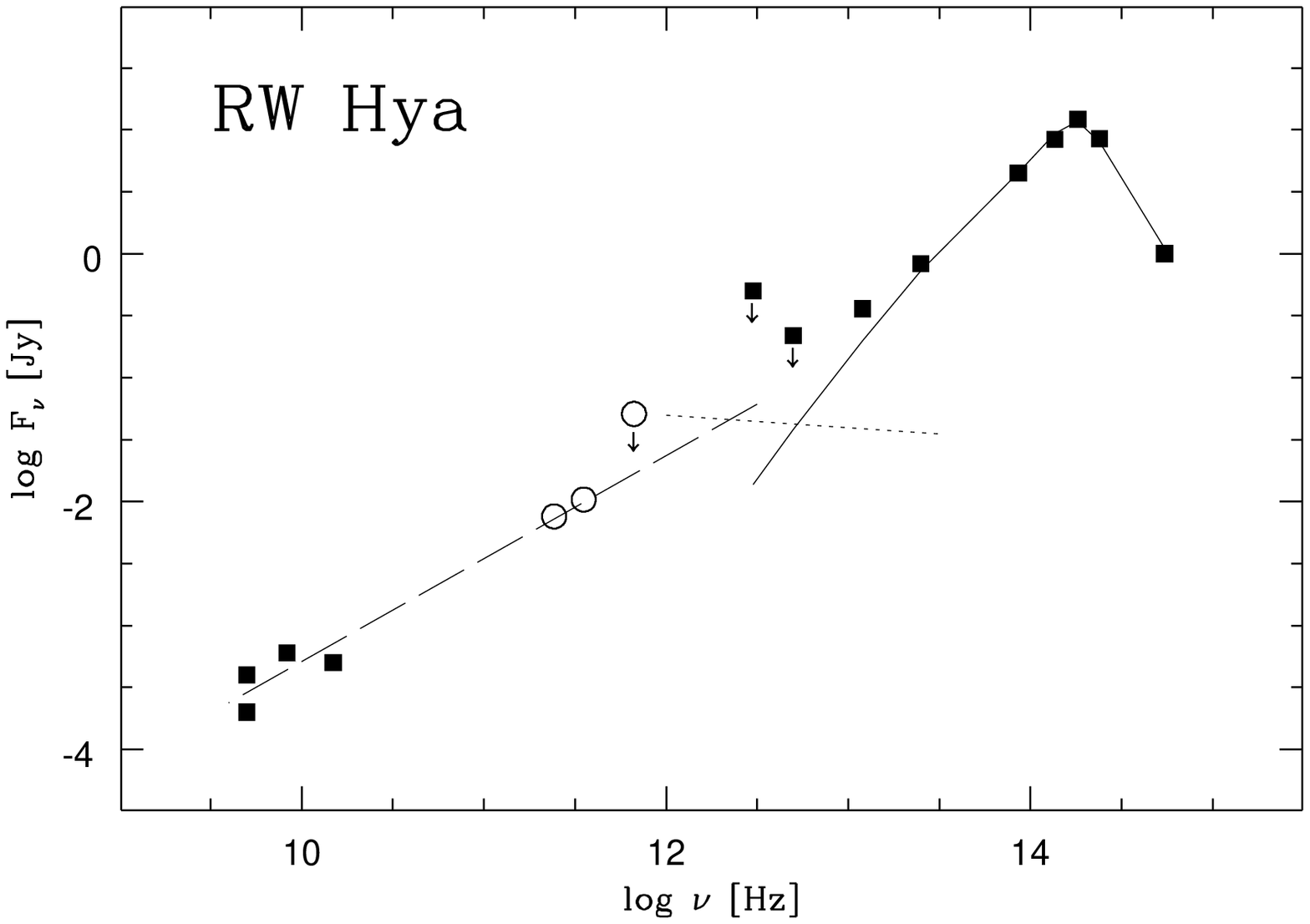}
\par\vspace{0pt}
\caption{\label{fig:fig3} Continuum spectrum of RW Hya covering the
radio, IR and visual bands. The spectrum is split into optically thick/thin
{\it ff} emission (dashed and dotted curves, respectively) and the giant
photospheric (solid curve) components.} 
\end{minipage}
\hfill
\begin{minipage}[t]{80mm}
\includegraphics[width=80mm]{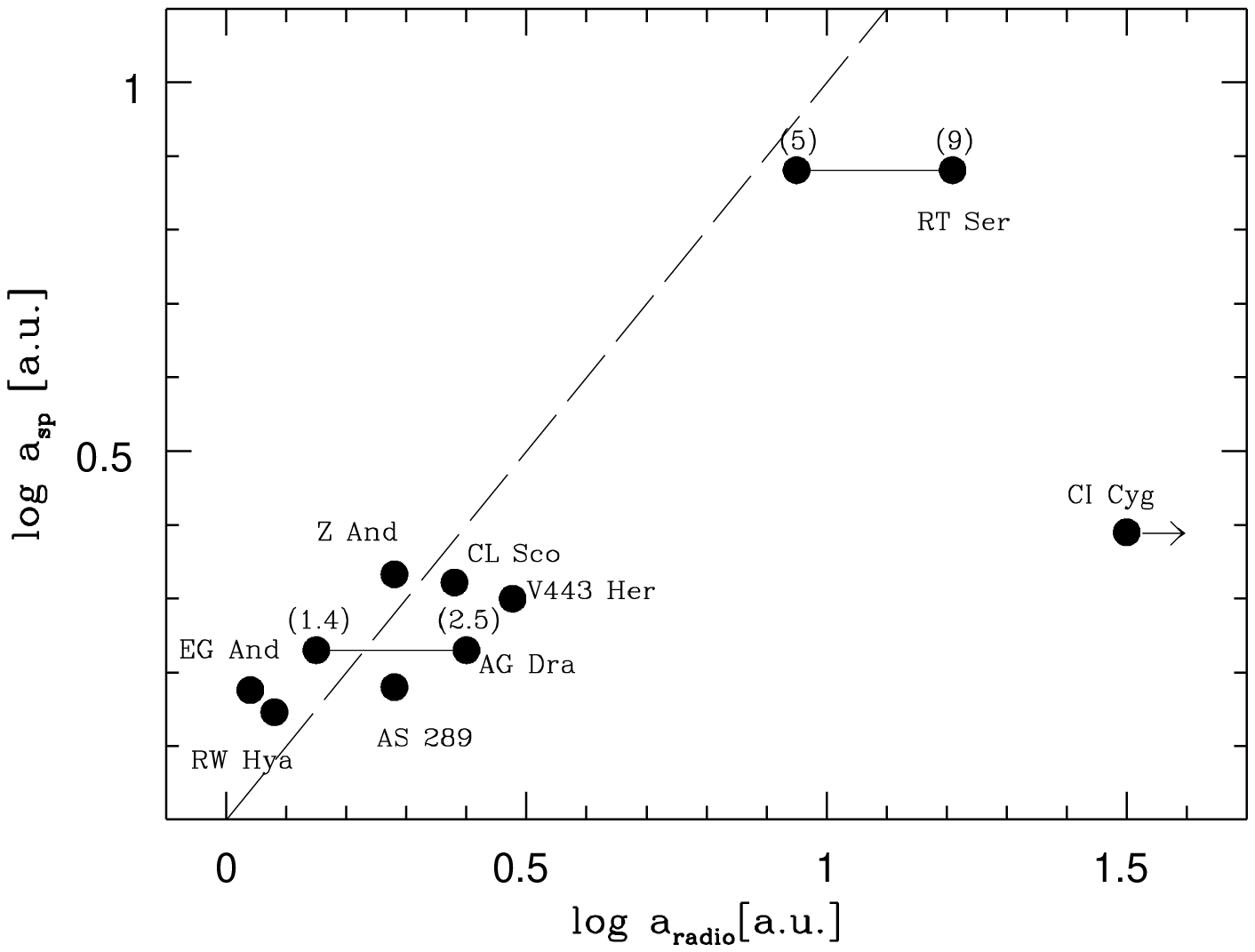}
\par\vspace{0pt}
\caption{\label{fig:fig.4} Comparison of the binary separations derived
from the radio data, $a_{\rm radio}$,
with independently known values, $a_{\rm sp}$.
For AG Dra and RT Ser two values of distance have been considered.}
\end{minipage}
\end{center}
\end{figure}

\section*{TESTING THE STB MODEL}

Figure 1 shows the relation between the radio flux density at 1.3 mm
plotted vs. the H$\beta$ flux corrected for interstellar extinction.
The diagonal lines represent optically thin case B emission for electron
temperatures appropriate for symbiotic nebulae. The systematic 
downward displacement with respect to the band shown suggests that the
radio emission is optically thick at least up to $\sim 1.3$ mm (243 GHz)
in agreement with the STB model which predicts the spectral turnover 
at $\sim 10^3$ GHz, i.e. at submm wavelengths, for typical quiescent S-type systems (SKT93). 
The mm radio emission shows also some correlation with the mid-IR flux,
and the radio luminosity increases with the $K-[12]$ colour (Figure 2), which
indicates that both the ionised gas and warm dust are involved in the
mass flow, and suggest that the cool giant may be the source of 
this material.
Similar results were obtained by SKT93 based on cm radio observations.

In fact, all stars in our sample but CI Cyg  have continua that rise with increasing frequency in  cm--mm range, and our IRAM and JMCT data
together with other radio data show a single power law with 
slope $\geq +0.6$ over nearly 3 decades in frequency. 
In Figure~3, for example, we show the spectrum of RW~Hya, which is typical
for non-variable S-type systems, and eruptive systems in quiescence. 
To constrain the turnover frequency, $\nu_{\rm t}$, for RW~Hya,
and other well-studied systems, we have estimated the
optically thin {\it ff} emission in the mm range using the H\,{\sc I}
{\it ff+bf} emission measure derived from the optical and UV data. 

In the STB model, knowledge of $\nu_{\rm t}$ allows
to estimate the binary separation, $a$, which can then
be compared with the independently known values.  
For a range in $X$ covering two orders of magnitude,
the binary separation within a factor of 2 is (ST90; Miko{\l}ajewska \&
Ivison 2001)
\begin{equation} 
a = 300\, (T_{\rm e}/10^4 {\rm K})^{-1/2} (\nu_{\rm t}/{\rm GHz})^{-1} 
(S_{\rm t}/{\rm mJy})^{1/2} (d/\rm kpc)\, \rm a.u.
\end{equation}
where $S_{\rm t}$ is the optically thin flux near the turnover,
and d is the distance.

The comparison of binary separations predicted by Eq.(1) is shown in Figure 4, 
and it provides the strongest support for the STB model.

The only system for which the STB model does not work is CI Cyg, which shows 
flat continuum in mm/sub-mm range with a turnover at $\sim 27$ GHz. Both the 
relatively low turnover frequency (as compared with other quiet S-type systems 
in our sample) and optically thin flux density require parameters entirely 
inconsistent with the well-known binary parameters of CI Cyg (Miko{\l}ajewska 
\& Ivison 2001). One possible cause of its low, optically thin radio emission 
(as compared with its very strong H\,{\sc i} {\it bf+ff} and line emission in 
the optical and near UV) is that CI Cyg is one of the few symbiotic 
systems, in which the M giant shows strong tidal distortion, and losses mass 
via Roche-lobe overflow rather than via stellar wind. 

\section*{IONIZATION GEOMETRY}

\noindent Most systems in our sample have radio spectra with the optically 
thick spectral index $\alpha$ greater than  0.6, which implies an ionization 
geometry with $X < \pi/4$, and for these cases knowledge of both the turnover 
frequency, $\nu_{\rm t}$, and $\alpha$ provides a measure of the value of $X$ 
which is related to the physical parameters of the system by the expresion 
(ST90, STB) 
\begin{equation} 
X = 4.7 \times 10^{-17} (a/{\rm  a.u.}) (L_{\rm 
ph}/10^{46}) [({\dot{M}/v})/({\rm M_{\odot} yr^{-1}/km\, s^{-1}})]^{-2} 
\end{equation} 
where $L_{\rm ph}$ is the Lyman continuum photon luminosity of 
the hot component, while $\dot{M}$ and $v$ are the mass-loss rate the terminal 
wind velocity of the cool giant, respectively. The spectral index never exceeds 
0.8, which implies $X \geq 0.2$ (STB).

This is consistent with results of numerical simulations of Raman scattered 
O\,{\sc vi} $\lambda\lambda$ 6825, 7082 emission lines observed in many 
symbiotic stars which suggest that symbiotic systems have preferentially an 
ionization geometry with an $X$-parameter of about 1, which means that the 
'average shape' of the ionization front does not differ significantly from the 
plane between the two stellar components (Schmid 1996). Moreover, the observed 
properties of the Raman scattered O\,{\sc vi} lines in AG Dra, Z And, and few 
other systems suggest scattering geometry with $0.4 \geq X \geq 4$ (Schmid \& 
Schild 1997a,b). Finally, recent study of optical spectra of 67 symbiotic 
systems shows that more than 50\%  of them have the O\,{\sc vi} 6825 
feature (Miko{\l}ajewska et al. 1997). The hot component luminosity in these 
systems is correlated with the luminosity in the O\,{\sc vi} 6825 line, which 
can be qualitatively understood in terms of similar scattering geometry, and so 
similar efficiency of the scattering process, for all systems that show this 
feature. In our sample, 7 objects show the O\,{\sc vi} lines, and it seems 
reasonable to assume that all of them have, within a factor of 4, $X \sim 1$.

\begin{figure}[h]
\begin{center}
\begin{minipage}[t]{80mm}
\includegraphics[width=80mm]{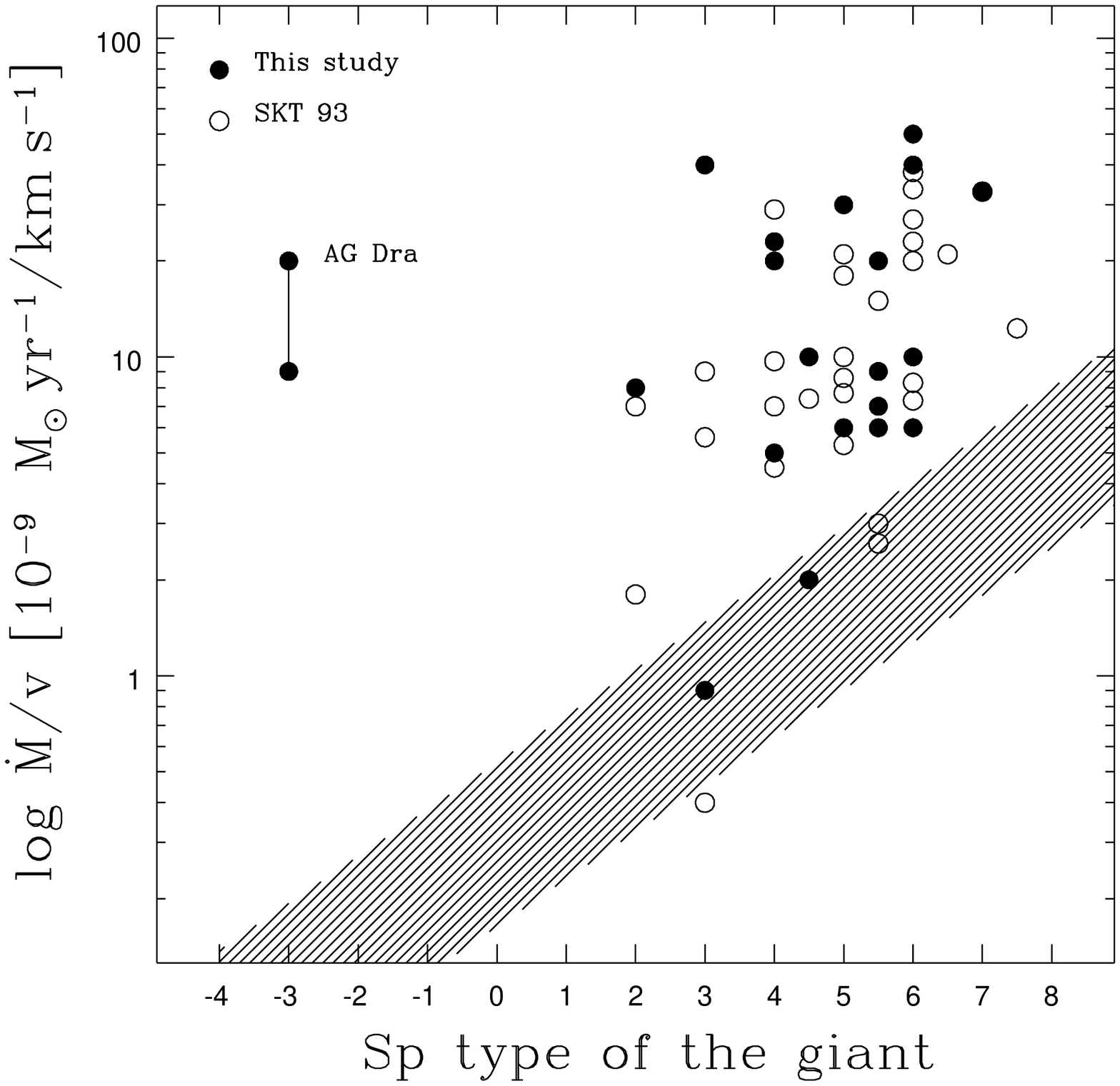}
\par\vspace{0pt}
\caption{\label{fig:fig5} Comparison of $\dot{M}/v$ estimates for symbiotic 
binaries with those for single field giants (shaded area).}
\end{minipage}
\hfill
\begin{minipage}[t]{80mm}
\includegraphics[width=80mm]{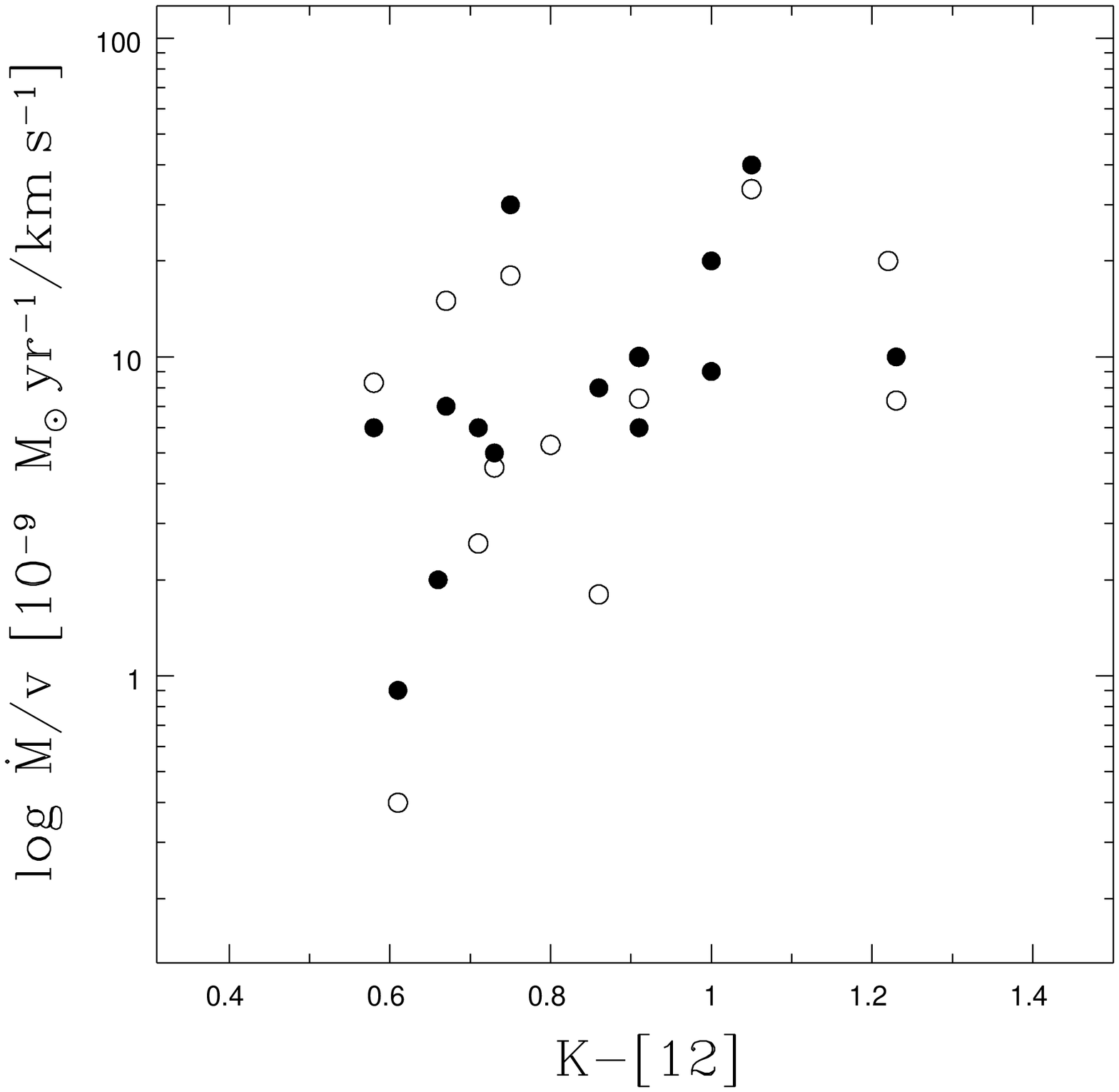}
\par\vspace{0pt}
\caption{\label{fig:fig.6} $\dot{M}/v$ vs. $K-[12]$ colour (same symbols
as in Fig.5).}
\end{minipage}
\end{center}
\end{figure}

\section*{MASS-LOSS RATES}

To estimate mass-loss rates for symbiotic giants from radio data, Wright \& 
Barlow's (1975) relation is usually applied. This relation, however, 
underestimates $\dot{M}$ if the wind is only partially ionized, which seems to 
be the case for most if not all quiescent symbiotic systems. SKT93 estimated 
the magnitude of this underestimate from numerical models as a factor of 2, 
1.5 and 1.15 for $X = 0.5, 1,$ and 5, respectively. Comparison of our estimates 
with $\dot{M}$'s for field giants (e.g. Dupree 1986) shows that symbiotic 
giants have systematically higher mass-loss rates than normal red giants do, 
and $\dot{M}/v$ is only weakly correlated with the spectral type of the giant 
(Figure 5).

Similar conclusion was reached by SKT93 based on analysis of radio emission at 
3.6 cm, and Kenyon et al. (1988), who found that many symbiotic giants have  
large $[25]-[12]$ and $K-[12]$ colour excesses that are not caused by 
interstellar extinction. Figure 6 shows that our mass-loss rates are correlated 
with $K- [12]$ colour. 

There are at least two possible explanation for the higher than average mass-
loss rates for symbiotic giants. First, it can be a selection effect in the 
sense that only the very evolved giants, and so those with the highest mass-
loss rates, can support symbiotic behaviour in widely separated binary systems. 
Second, it is possible that the mass-loss rate of the symbiotic giant is 
enhanced by its binary companion, which reduces the gravity at some points in 
the outflowing material due to illumination  heating and/or causes tidal 
friction and perhaps enhanced dynamo activity. Tout \& Eggleton (1988) argued 
that the tidal friction enhances mass loss by up to $\sim 2$ orders of 
magnitude in binary systems containing red giants. Tidal interactions are 
certainly important in symbiotic systems as suggested by practically circular 
orbits of all our target sample, as well as of most ($\sim 80$\%) other 
symbiotic systems with known orbital solutions (Belczy{\'n}ski et al. 2000).

On the other hand, Figure 7 shows that the giant mass-loss rate is apparently 
correlated with the hot component luminosity, $L_{\rm h}$, which suggests 
importance of the illumination heating of the outer atmosphere of the red 
giant. Although one can naively interpret this correlation as a proof that in 
fact the radio emission originates from a wind emanating predominantly from the 
hot component, this is certainly not the case for our targets. The hot 
component wind with $\dot{M}/v \sim 10^{-8} \rm M_{\odot}\,yr^{-1}$ should 
produce prominent P-Cyg profiles in high-excitation emission lines which are 
not observed. Moreover, in such case the turnover frequency would be not 
related to the binary separation. On the contrary, the very good agreement 
between the binary separations derived from $\nu_{\rm t}$ and the spectroscopic 
orbits (Figure 4) as well the correlation of $\dot{M}/v$ with $K-[12]$ colour 
provide strong evidence that the wind originates from the cool giant.

\begin{figure}[h]
\begin{center}
\begin{minipage}[t]{80mm}
\includegraphics[width=80mm]{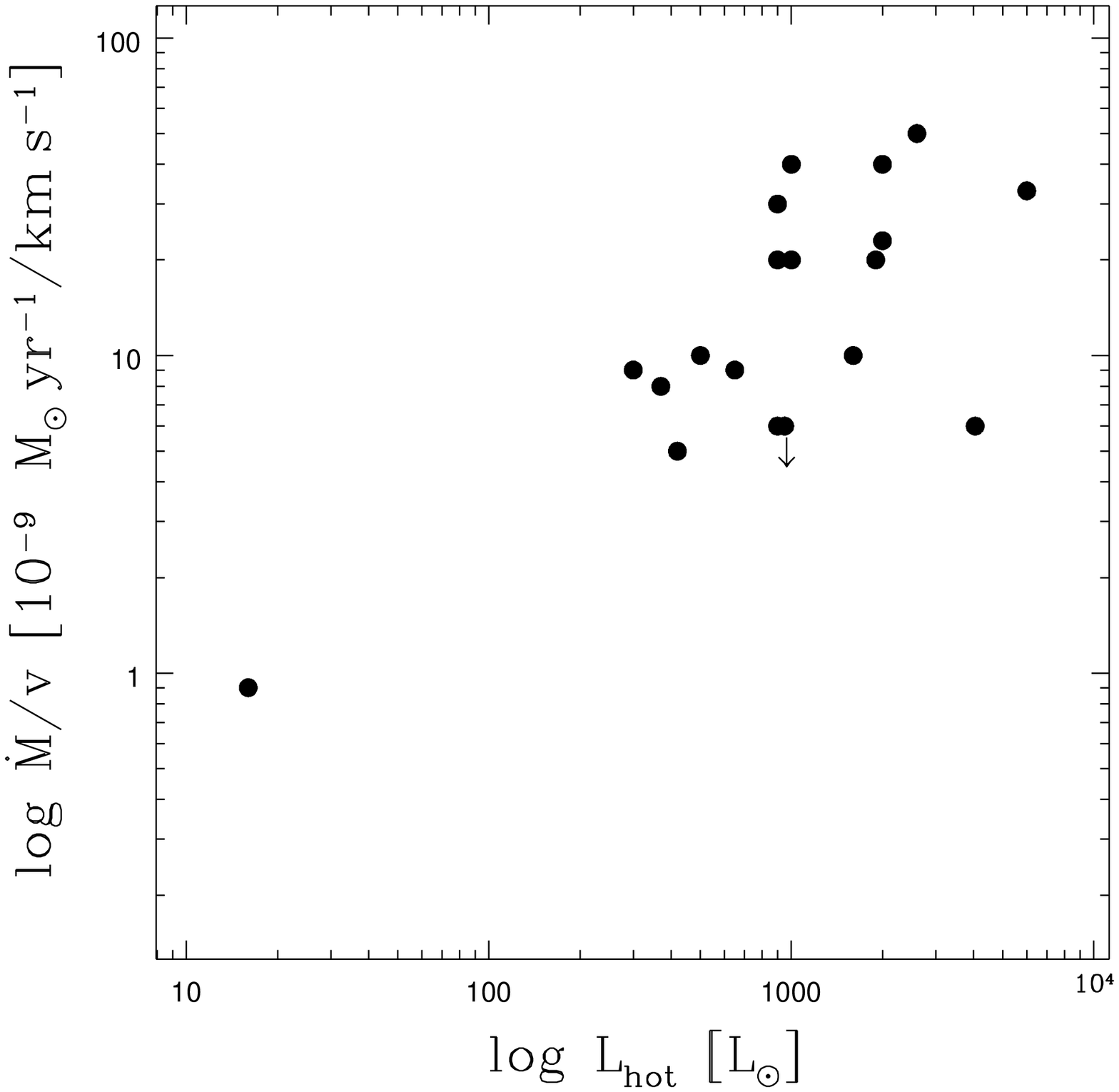}
\par\vspace{0pt}
\caption{\label{fig:fig7} The cool giant mass-loss rate vs. the
hot component luminosity.}
\end{minipage}
\hfill
\begin{minipage}[t]{80mm}
\includegraphics[width=80mm]{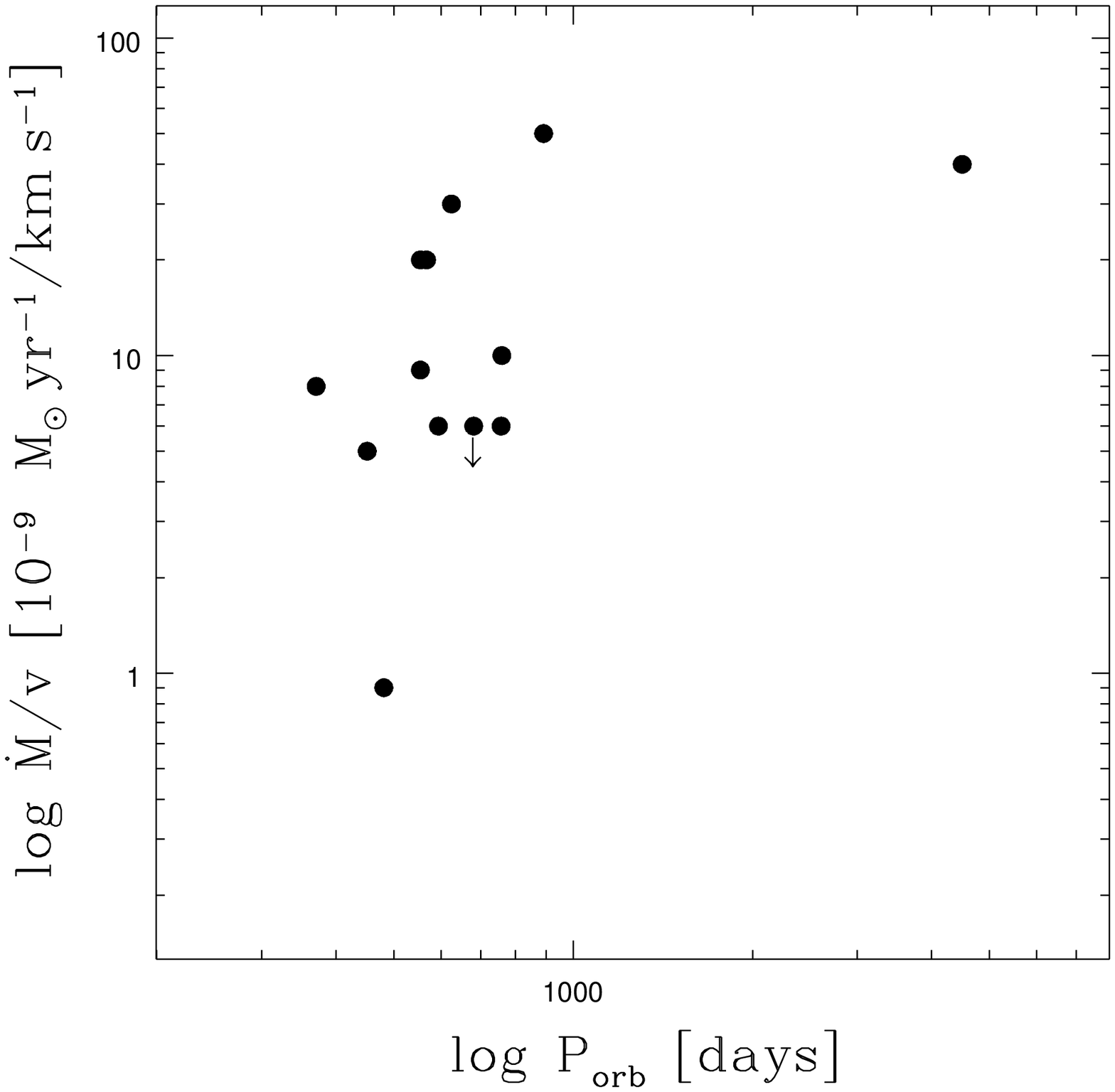}
\par\vspace{0pt}
\caption{\label{fig:fig.8} The cool giant mass-loss rate vs. 
binary period.}
\end{minipage}
\end{center}
\end{figure}

The correlation between the giant mass-loss rate and $L_{\rm h}$ also has a 
natural interpretation in the frame of proposed models for the hot component. 
In general, the stronger is the cool giant wind the more material can be 
accreted by its hot companion. Most symbiotics interact by wind-driven 
accretion, and the expected accretion rate is a few per cent of $\dot{M}_{\rm 
wind}$, and in the case of our sample targets the corresponding accretion 
luminosity is of order of $10-100\, \rm L_{\odot}$. Although the hot component 
of a typical quiescent symbiotic system, with $L_{\rm h} \sim 1000 \rm 
L_{\odot}$ and $T \sim 10^5$~K, cannot be powered solely by accretion, the 
situation improves radically if the white dwarfs burn H-rich material as they 
accrete it. For a typical $0.4-0.6\, \rm M_{\odot}$ symbiotic white dwarf 
(Miko{\l}ajewska 1997) the minimum accretion rate at which this happens is of 
order of $10^{-8} \rm M_{\odot}\,yr^{-1}$, while the maximum steady burning 
rate is set by the core mass-luminosity relation, and it is a few times higher. 
Thus depending on accretion rate (which is related to the giant mass-loss rate) 
the hot component luminosity is powered either by accretion or by thermonuclear 
burning of the accreted material, and in both cases its luminosity will be 
somehow related to the giant mass-loss rate. The mass-loss rates for symbiotic 
giants derived from our radio observations are thus sufficient to power via 
wind-accretion and thermonuclear burning the observed luminosities of their hot 
companions.

Finally, the giant-mass loss rates tend to increase with the orbital period 
(Figure 8). It is interesting that a correlation between the hot component 
luminosity and the orbital period was first find by Miko{\l}ajewska \& Kenyon 
(1992), while M{\"u}rset \& Schmid (1999) found a relation between the giant 
spectral type and the orbital period. Miko{\l}ajewska \& Kenyon suggested that 
their result in fact implies a relation between the white dwarf mass and the 
binary period  if the hot components lie on the plateau portion of white dwarf 
cooling curves and follow some standard core-mass luminosity relation. 
M{\"u}rset \& Schmid noted that the limiting line from the spectral type -- 
orbital period diagram is practically identical with the relation $l_1=2R_{\rm 
giant}$ (where $l_1$ is the distance from the giant's center to the Lagrangian 
point $\rm L_1$), suggesting that this configuration is ideal for producing 
long-lived symbiotic phenomena. 

\section*{CONCLUSIONS}

We conclude that the radio emission in quiescent S-type systems
except CI~Cyg (and other systems containing Roche-lobe filling giants)
originates from  the red giant wind partially ionized by the hot component
as proposed by STB.
The mass-loss rates from  symbiotic giants are systematically higher
that in single field giants which suggest that only the giants with the 
highest mass-loss rates can support symbiotic behaviour in widely
separated binary systems, in particular can power via wind-accretion 
and subsequent thermonuclear burning the observed luminosities 
of  their hot companions.

\section*{ACKNOWLEDGEMENTS}
This research has been supported by the KBN grant 2P03D02112,
and by the JUMELAGE program ``Astronomie France-Pologne''
of CNRS/PAN. RJI acknowledge the award of a PPARC Advanced Fellowship.

\end{document}